\documentclass[a4paper]{article}

\usepackage{INTERSPEECH2020}
\usepackage{multirow}
\newcommand{\tabincell}[2]{\begin{tabular}{@{}#1@{}}#2\end{tabular}}

\title{A Comparison of Label-Synchronous and Frame-Synchronous End-to-End Models for Speech Recognition}
\name{Linhao Dong$^{1,2,*}$, Cheng Yi$^{1,2,*}$, Jianzong Wang$^{3, \dagger}$, Shiyu Zhou$^1$, Shuang Xu$^1$, Xueli Jia$^{3}$, Bo Xu$^1$
\thanks{* denotes equal contribution to this work}
\thanks{$\dagger$ denotes the corresponding author}
}
\address{
$^1$Institute of Automation, Chinese Academy of Sciences, China\\
$^2$University of Chinese Academy of Sciences, China \\
$^3$Ping An Technology (Shenzhen) Co., Ltd.}
\email{\{donglinhao2015, yicheng2016, zhoushiyu2013, shuang.xu, xubo\}@ia.ac.cn \\ \{wangjianzong347, jiaxueli373\}@pingan.com.cn}

\begin{document}
\maketitle
\begin{abstract}
End-to-end models are gaining wider attention in the field of automatic speech recognition (ASR). One of their advantages is the simplicity of building that directly recognizes the speech frame sequence into the text label sequence by neural networks. According to the driving end in the recognition process, end-to-end ASR models could be categorized into two types: label-synchronous and frame-synchronous, each of which has unique model behaviour and characteristic. In this work, we make a detailed comparison on a representative label-synchronous model (transformer) and a soft frame-synchronous model (continuous integrate-and-fire (CIF) based model). The results on three public dataset and a large-scale dataset with 12000 hours of training data show that the two types of models have respective advantages that are consistent with their synchronous mode.
\end{abstract}
\noindent\textbf{Index Terms}: End-to-end, ASR, frame-synchronous, label-synchronous

\section{Introduction}
End-to-end models are profoundly affecting the development of automatic speech recognition (ASR), and have demonstrated their advantages on a wide range of ASR tasks. Since end-to-end ASR models directly transform the speech sequence (usually in the form of feature frames) into the text label sequence by neural networks, there must be one end (the text end or the speech end) to drive the recognition process of the entire model. According to the driving end, current end-to-end models could be categorized into two types: label-synchronous models and frame-synchronous models.

Label-synchronous models refer to the models that are driven by the text end: they process one label at each step and stop after the label of end of sentence is recognized. The representative models in them are originally proposed for neural machine translation (NMT) \cite{bahdanau2015neural, vaswani2017attention}, and rely on the attention mechanism to extract relevant speech information from the encoded frames, such as the attention-based models \cite{chorowski2015attention, chan2016listen, jaitly2015neural, raffel2017online, chiu2017monotonic, hou2017gaussian} and transformer \cite{dong2018speech, karita2019comparative, miao2020transformer}. Compared with frame-synchronous models, they have two main advantages: 1) since they usually attend to a large range of encoded frames, they could utilize the speech information more comprehensively; 2) since they are label-synchronous, they could jointly extract the relevant information for each label from multiple channels \cite{chang2019mimo} or sources \cite{pundak2018deep, bruguier2019phoebe}. Recently, since transformer has shown performance superiority in the comparison with the RNN attention-based model \cite{karita2019comparative}, we focus on the model of transformer in this work.

\begin{figure}
  \centering
  \includegraphics[width=0.85\linewidth]{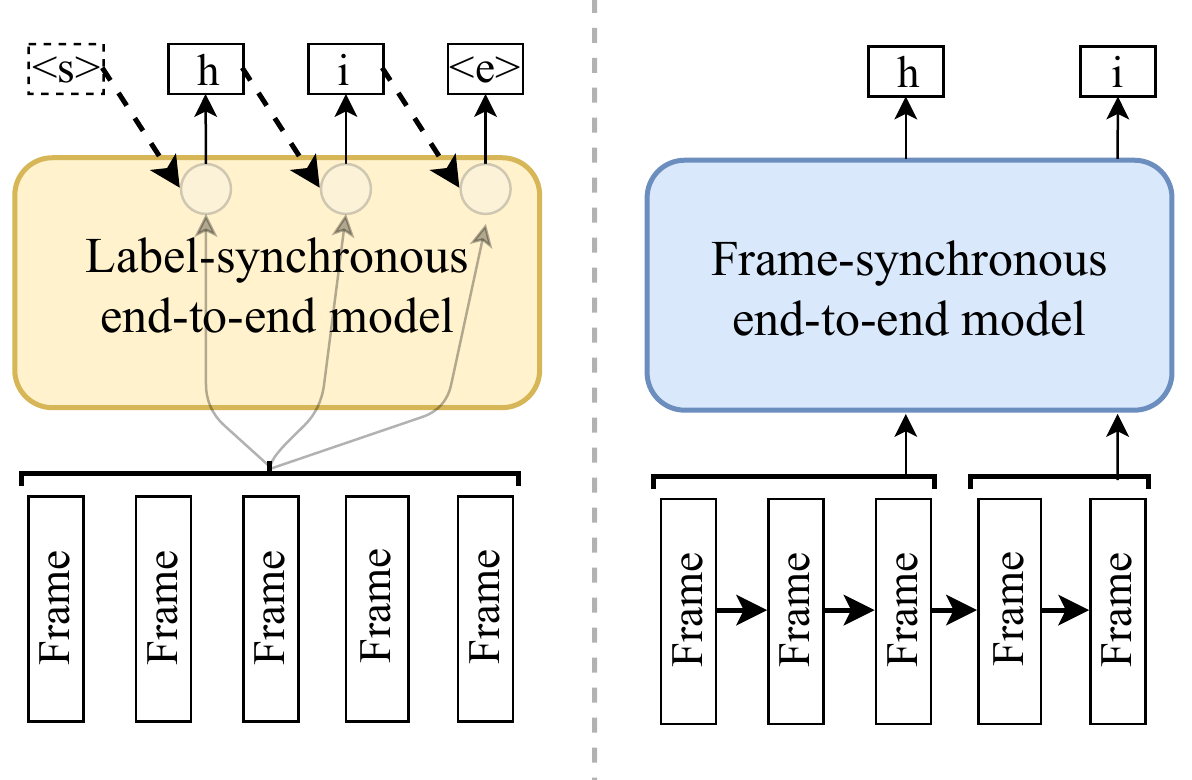}
  \vspace{-4mm}
  \caption{Schematic diagram of the label-synchronous end-to-end model and the soft frame-synchronous end-to-end model.}
  \label{fig:simplified_figure}
  \vspace{-6mm}
\end{figure}

Frame-synchronous models refer to the models that are driven by the speech end: they process one frame at each step and stop after the last frame is processed. Compared with label-synchronous models, they have two main advantages: 1) since they process in the frame-by-frame manner, they could naturally support online speech recognition; 2) since they are aware of the frame position, they could provide time stamp for the recognition result. The representative models in them could be categorized into two types: 1) the models that use a hard \cite{battenberg2017exploring} alignment, including the CTC \cite{graves2006connectionist} and the CTC-like \cite{graves2012sequence, sak2017recurrent} models; 2) the models that use a soft \cite{battenberg2017exploring} alignment, which first locates the acoustic boundary through the frame-by-frame detection, and then integrate the acoustic information from the located speech piece using the form of weighted-sum (e.g. \cite{moritz2019triggered, li2019end, dong2020cif}). Compared with the hard ones, the soft frame-synchronous models exclude the influence of the blank label, thus having less decoding computation and simpler search process. In this work, we focus on the soft frame-synchronous models and experiment on the continuous integrate-and-fire (CIF) based model \cite{dong2020cif}, which is calculated in a concise manner and uses the same self-attention network (SAN) as transformer.

The concepts of label-synchronous and frame-synchronous first appear in \cite{prabhavalkar2017comparison}, where the authors compare the attention-based model \cite{chan2016listen} with the CTC \cite{graves2006connectionist} and RNN-T \cite{graves2012sequence} models in detail. Differing from it: 1) we think the concepts of label-synchronous and frame-synchronous should not only be used to describe the decoding manner \cite{prabhavalkar2017comparison}, but could be used to describe end-to-end models with different model behaviour and characteristic. By distinguishing the two types of models, it not only benefits to clarify the advantages and disadvantages of them, but also could provide guidance for applying the suitable model in specific ASR scenarios; 2) we focus on the comparison of a label-synchronous model (transformer) with a soft frame-synchronous model (CIF-based model), rather than the hard frame-synchronous models in \cite{prabhavalkar2017comparison}. Specifically, we make a detailed comparison for the two models on multiple datasets, including three public sets and a large-scale data set with 12000 hours of training data. We find the two models show their respective advantages on accuracy, speed and generalization. The detailed contents will be introduced in the following sections.

\section{Models}
In this section, we describe the label-synchronous model (transformer) and the frame-synchronous model (continuous integrate-and-fire (CIF) based model) compared in this work. Both of them follow the encoder-decoder framework, where the encoder transforms the speech feature sequence $x = \{x_1, ..., x_T\}$ into the encoded acoustic representations $h = \{h_1, ..., h_U\}$, the decoder receives the previous label $y_{i-1}$ and the relevant acoustic information to predict the current label $y_i$. The encoder and the decoder are connected by the alignment mechanism, which determines the extraction manner of the relevant acoustic information used for decoding. Next, we will combine the diagram of the alignment mechanism (as shown in figure \ref{fig:alignment_figure}) to introduce the two models.

\subsection{Transformer}
The transformer model used in this work has the similar structure as \cite{dong2018speech, karita2019comparative}, but has two differences: 1) it abandons the sinusoidal positional encoding in \cite{dong2018speech, karita2019comparative} and uses the proximity bias in the self-attention network (SAN) \cite{dong2019self} to provide relative positional information (since this performs better and more stable in our experiments); 2) it uses the encoder structure in \cite{dong2019self}, which uses a convolutional front-end and a pyramid of SAN.

The running of transformer is driven by the text end, or the decoder end. At each decoder step $i$, the previous label $y_{i-1}$ is input to a stack of $N_d$ identical decoder blocks, each of which is composed by inserting a multi-head encoder-decoder attention into the two sub-networks (multi-head self-attention and position-wise feed-forward network) of the SAN. Each of the multi-head encoder-decoder attention receives the encoded representation $h$ as the key and value, and receives the outputs of the multi-head self-attention (in the same decoder block) as the query, then they conduct the multi-head attention to extract the relevant acoustic information. As shown in figure \ref{fig:alignment_figure} (a), transformer runs label by label, and stops after the label of end of sentence ($\langle$ e $\rangle$) is predicted. Assume the length of the predicted labels is $S$, the length of $h$ is $U$, the computational complexity of the alignment mechanism in transformer is $O(N_d \cdot U \cdot S)$.

\subsection{CIF-based model}
The CIF-based model used in this work follows the model structure in \cite{dong2020cif}. Its encoder first generates the encoded representation $h_u$ for each encoder step $u$. Then, it predicts a weight $\alpha_u$ for each $h_u$ by passing a window centered at $h_u$ (e.g. $[h_{u-1},h_u,h_{u+1}]$) to a 1-dimensional convolutional layer and then a fully connected layer with one output unit and a sigmoid activation. Next, it passes the weight $\alpha_u$ and $h_u$ to CIF, which forwardly accumulates the weight and integrates the acoustic representation (using the form of weighted sum) until the accumulated weight reaches a threshold (1.0), which means an acoustic boundary is located. At this point, CIF divides current weight $\alpha_u$ into two part: the one is used to fulfill the integration of current label $y_i$ by building a complete distribution (whose sum of weights is 1.0) on relevant encoder steps, the other is used for the integration of next label. After that, it fires the integrated acoustic embedding $c_i$ (as well as the context vector) to the decoder to predict the corresponding label $y_i$.

\begin{figure}
  \centering
  \includegraphics[width=\linewidth]{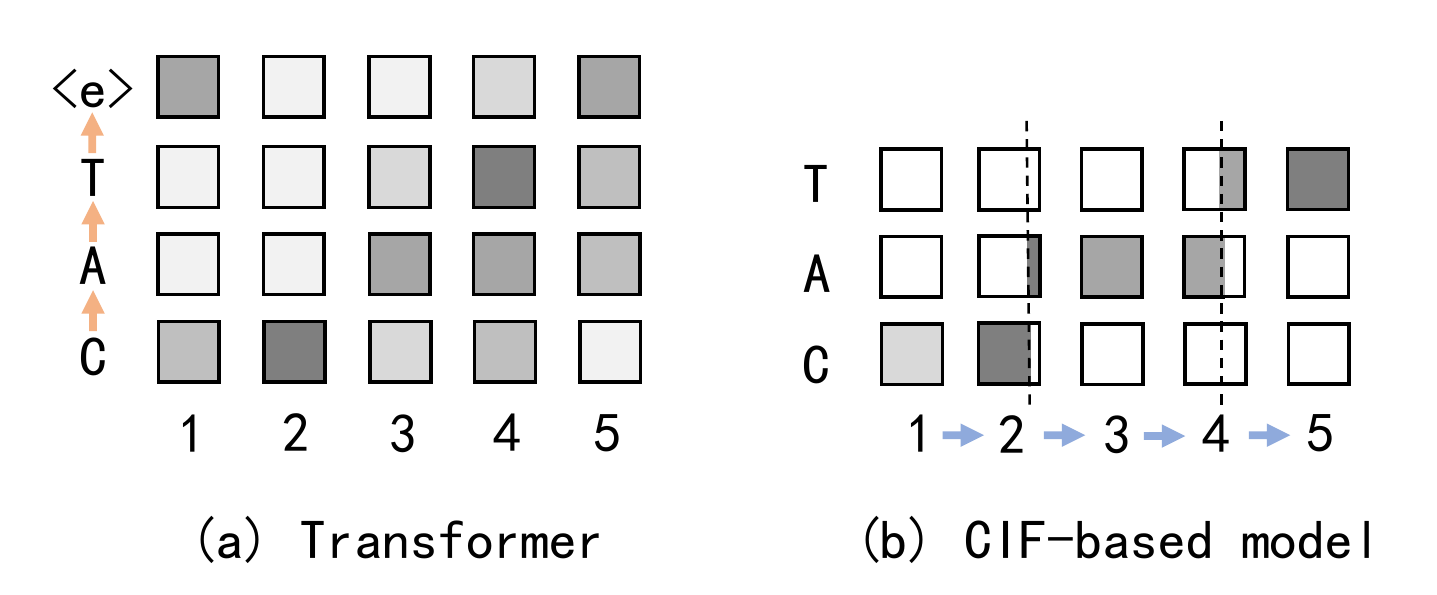}
  \vspace{-4mm}
  \caption{Schematic diagram of the alignment mechanism used in transformer and the CIF-based model when processing on an utterance with 5 frames and is labelled as "CAT". The shade of gray in each square represents the weight of each encoder step involved in the calculation of decoding labels. The arrows indicate the synchronous mode of the two models.}
  \label{fig:alignment_figure}
  \vspace{-6mm}
\end{figure}

The running of CIF-based model is driven by the speech end, or the encoder end. At each encoder step $u$, the representation $h_u$ and the weight $\alpha_u$ of the encoded frame $u$ are input to the CIF module, which determines if there have integrated enough acoustic information to trigger the calculation of the decoder. As shown in figure \ref{fig:alignment_figure} (b), the CIF-based model runs frame by frame, and stops after the last frame is processed. Assume the length of $h$ is $U$, the computational complexity of the CIF alignment mechanism is $O(U)$.

\subsection{Model details}
The two models use the same encoder structure \cite{dong2020cif}, which reduces the frame rate to 1/8. The CIF-based model uses the autoregressive decoder that is based on SAN. The decoder of both the two models caches the computed SAN states of all heads for efficient inference. In the training, both of the two models use multi-task learning: the CIF-based model is trained under the cross-entropy loss, the CTC loss on the encoder with coefficient $\lambda_1$ and the quantity loss on the CIF with coefficient $\lambda_2$ \cite{dong2020cif}, and the transformer is trained under the cross-entropy loss and the CTC loss on the encoder \cite{karita2019comparative}. In the inference, we first perform beam search to get the decoded hypothesis, then use the score of a SAN-based language model (LM) with coefficient $\gamma$ to rescore the hypothesis predicted by the two models.

\section{Experimental Setup}
\label{sec:exp_setup}

\begin{table*}[!ht]
\centering
\caption{Results of accuracy performance for the models compared in this work. The evaluation metric uses word error rate (WER) for Librispeech, and uses character error rate (CER) for the three Mandarin datasets, and keep the same for the later table \ref{tab:res_long} and \ref{tab:res_repeat_noise}.}
\vspace{-2mm}
\setlength{\tabcolsep}{1.5mm}{
\begin{tabular}{c|c|c|c|c|c|c|c|c|c}
\toprule
\multirow{2}{*}{Model} & \multicolumn{2}{c|}{Librispeech} & \multicolumn{3}{c|}{AISHELL-2} & \multirow{2}{*}{HKUST} & \multicolumn{3}{c}{CN12000h} \\
\cline{2-3} \cline{4-6} \cline{8-10}
& test\_clean & test\_other & test\_android & test\_ios & test\_mic & & test\_set 1 & test\_set 2 & test\_set 3 \\
\midrule
Transformer & \textbf{2.78} & \textbf{7.57} & 6.12 & 5.74 & 6.31 & \textbf{21.88} & \textbf{5.52} & \textbf{4.61} & \textbf{22.27} \\
CIF-based model & 2.86 & 8.08 & \textbf{6.09} & \textbf{5.68} & \textbf{6.20} & 22.80 & 5.54 & 5.01 & 23.01 \\
\bottomrule
\end{tabular}
}
\vspace{-5mm}
\label{tab:res_accuracy}
\end{table*}

We conduct the comparison on three public ASR datasets and a large Mandarin ASR dataset (CN12000h) that contains about 12000 hours of training data. The three public ASR datasets include the English read-speech dataset (Librispeech \cite{panayotov2015librispeech}) that contains 960 hours of training data, the Mandarin read-speech dataset (AISHELL-2 \cite{du2018aishell}) that contains 1000 hours of training data, and the Mandarin telephone ASR benchmark (HKUST \cite{liu2006hkust}) that contains 168 hours of training data. Other details about the usage of three datasets are introduced in \cite{dong2020cif}. For the CN12000h dataset, the training set covers data from different application scenarios, so we use three representative test sets to evaluate the performance of the models and use one of them (test\_set 2) as the development set.

For the three public ASR datasets, the generation of input features and output labels keeps the same as \cite{dong2020cif}. The frequency masking and time masking in SpecAugment \cite{park2019specaugment} with $F=8$, $m_F=2$, $T=70$, $m_T=2$, $p=0.2$ are applied to the models on the three datasets. For the CN12000h dataset, the input features use 29 dimensional filterbanks extracted from 25 ms window and shifted every 10 ms, then extended with delta and delta-delta, and the global normalization. The output labels cover 4733 classes including 4701 Chinese characters, 26 uppercase English letters, 3 special marker (noise, etc), the blank, the label of end-of-sentence and the pad. Above setting keeps the same for the two models compared in this work.

For the CIF-based model, the encoder uses the 2-layer convolutional front-end and the pyramid of self-attention network (SAN) in \cite{dong2019self}, where the head number $h$, the hidden size $h_{model}$ and the inner size $h_{ff}$ in SAN are set to $4, 640, 2560$ for all Mandarin datasets and set to $4, 1024, 4096$ for Librispeech, $n$ in the pyramid structure is set to 5 for all datasets, thus the number of SAN encoder layers is 15. The 1-dimensional convolutional layer that predicts weights uses $d_{model}$ convolutional filters, and the convolutional width is 3. Besides, layer normalization \cite{ba2016layer} and a ReLU activation are applied after this convolution. The decoder uses the autoregressive decoder in \cite{dong2020cif} and the number of SAN decoder layers is 2. The coefficient $\lambda_1$ on the CTC loss is set to 0.5 for all Mandarin datasets and is set to 0.25 for Librispeech, $\lambda_2$ on the quantity loss is set to 1.0.

For transformer, it uses the same $h$, $h_{model}$ and $h_{ff}$ as the SAN in the CIF-based model. The encoder uses the same structure as the CIF-based model except $n$ in the pyramid structure is set to 4 for all datasets, thus the number of SAN encoder layers is 12. The decoder uses 6-layer decoder blocks. The setting of the layer number in the encoder and the decoder refers to the best setting in \cite{dong2018speech, karita2019comparative}, and it also makes the two models have similar amount of parameters. The coefficient on the CTC loss keeps the same as the CIF-based model.

The language model (LM) uses SAN with $h=4$, $d_{model}=512$, $d_{ff}=2048$ for the three public datasets, and the number of SAN layers is set to 3, 6, 20, 6 for HKUST, AISHELL-2, Librispeech and CN12000h, respectively. Above models are implemented on TensorFlow \cite{abadi2016tensorflow}.

We use almost the same training and inference strategy for the two models. In the training, we batch the data with approximate frame length together and fill about 20000 frames for every batch of the three public dataset, and fill 37500 frames (the maximum for a p6000 gpu) for every batch in CN12000h. We use 4, 8, 8, 8 gpus to train the models on HKUST, AISHELL-2, Librispeech and CN12000h, respectively. We use the optimizer and the varied learning rate formula in \cite{dong2018speech}, where the warmup step is set to 25000 for HKUST and is set to 36000 for other datasets, the global coefficient is set to 4.0. We only apply dropout to the SAN, whose attention dropout and residual dropout are all set to 0.2 except the models on CN12000h that is set to 0.1. We use the uniform label smoothing in \cite{chorowski2017towards} and set it to 0.2 for both of the ASR models and the language models. Parallel scheduled sampling (PSS) \cite{duckworth2019parallel} with a constant sampling rate of 0.5 is applied to the models on the three Mandarin datasets. After training, we average the newest 10 checkpoints for inference. In the inference, we use beam search with size 10 for all models. For the CIF-based model, the coefficient $\gamma$ for LM rescoring is set to 0.1, 0.2, 0.9, 0.2 for HKUST, AISHELL-2, Librispeech and CN12000h respectively. For transformer, the hyper-parameter $\gamma$ is set to 0.6, 0.9, 2.0, 0.6 for HKUST, AISHELL-2, Librispeech and CN12000h, respectively.

\section{Results}
In this section, we present the results of the compared two models on the performance of accuracy, speed and generalization. The two models are evaluated on the offline mode, and use almost the same training strategy, similar model structure and similar amount of parameters to make the comparison as fair as possible, the details are introduced in section \ref{sec:exp_setup}.

\subsection{Comparison on accuracy performance}
As shown in table \ref{tab:res_accuracy}, we find transformer performs better on 3/4 datasets (Librispeech, HKUST and CN12000h), while the CIF-based model only shows better performance on AISHELL-2, which is a Mandarin read speech dataset that has clear acoustic boundary between labels. To our best knowledge, the result of CIF-based model on AISHELL-2 and the result of transformer on HKUST are the best accuracy performance of respective dataset, thus the comparison is conducted on two very strong models that have different synchronous mode.

From the perspective of synchronous mode, the accuracy difference between the two models can be attributed to two aspects: 1) since transformer is a label-synchronous model and extracts acoustic information of each label from the global view, it could make comprehensive use of the encoded acoustic information and capture more acoustic details for the decoding. In contrast, the CIF-based model performs a frame-by-frame manner and doesn't cache the processed frames like \cite{moritz2020streaming}, thus it integrates acoustic information from a limited local view, which may affect its modeling expressiveness to some extent; 2) since the CIF-based model is a soft frame-synchronous model that needs to locate the acoustic boundary during processing, this makes it perform slightly inferior on the datasets with blurred acoustic boundary between labels. In contrast, transformer is less affected by the clearness of acoustic boundary. In addition to the above, the multi-layer encoder-decoder attention of transformer (from its $N_d$ decoder blocks) also benefits to its extraction of acoustic information and promotes better accuracy.

\subsection{Comparison on speed performance}
\label{sec:res_speed}
The accuracy advantage of transformer is partly due to the global encoder-decoder attention. However, attending to every encoder step is bound to bring a mass of unnecessary calculations on steps that are acoustically irrelevant to the decoding label, which may affect its calculation speed. Thus we also compare the speed performance, including the training speed and the real time factor (RTF) (= inference time / audio time).

\begin{table}[!ht]
\centering
\caption{Results of speed performance for the models compared in this work. The following results are obtained on the Quadro P6000 gpu.}
\vspace{-2mm}
\begin{tabular}{c|c|c|c}
\toprule
Dataset & Model & \tabincell{c}{Training \\ speed \\ (step/sec)} & \tabincell{c}{Real Time \\ Factor \\ (RTF) } \\
\midrule
\multirow{2}*{Librispeech} &
Transformer & \textbf{0.407} & 0.0815 \\
& CIF-based model & 0.308 & \textbf{0.0120} \\
\midrule
\multirow{2}*{AISHELL-2} &
Transformer & \textbf{0.847} & 0.0209 \\
& CIF-based model & 0.630 & \textbf{0.0048} \\
\midrule
\multirow{2}*{HKUST} &
Transformer & \textbf{0.987} & 0.0226 \\
& CIF-based model & 0.873 & \textbf{0.0049} \\
\bottomrule
\end{tabular}
\vspace{-5mm}
\label{tab:res_speed}
\end{table}

As shown in table \ref{tab:res_speed}, we find transformer has faster training speed than the CIF-based model. Since both of them use the SAN-based encoder and decoder, which are computed in a highly parallel manner, the gap on the training speed can be attributed to the frame-by-frame incremental calculation by the CIF. Even so, since the incremental calculation by the CIF is lightweight, the training efficiency of the CIF-based is still considerable and is about $3/4 - 8/9$ of transformer.

Besides, we find the CIF-based model obtains about $1/6.8-1/4.4$ of the RTF than transformer, which means it obtains about $4.4-6.8$ times inference speed than transformer. Since the two models are compared on the offline mode, both of them perform one-time encoder calculation and multi-time decoder calculations, the heavy decoder of transformer that takes $O(N_d \cdot U \cdot S)$ cost to extract the acoustic information inevitably brings large amount of calculations. In contrast, the CIF-based model that takes $O(U)$ cost to integrate the acoustic information has much lighter inference calculation, which is one of the advantages of this type of frame-synchronous models.

\subsection{Comparison on generalization}
In addition to the accuracy and speed, we wonder to know whether the synchronous mode affects the modeling generalization on some special cases.

\begin{table}[!ht]
\vspace{-2mm}
\centering
\caption{Results on the long utterances with multiple durations. The `s' after the figures represents seconds.}
\vspace{-2mm}
\setlength{\tabcolsep}{0.5mm}{
\begin{tabular}{c|c|c|c|c|c|c}
\toprule
Dataset & Model & \tabincell{c}{20s-\\30s} & \tabincell{c}{30s-\\40s} & \tabincell{c}{40s-\\50s} & \tabincell{c}{50s-\\60s} & \tabincell{c}{60s-\\70s}\\
\midrule
\multirow{2}*{\tabincell{c}{Librispeech \\ (test-clean)}} &
Transformer & \textbf{3.64} & 4.12 & 3.68 & 6.40 & 11.51 \\
& CIF-based model & 4.14 & \textbf{3.72} & \textbf{2.68} & \textbf{3.38} & \textbf{3.61} \\
\midrule
\multirow{2}*{\tabincell{c}{AISHELL-2 \\ (test-ios)}} &
Transformer & 16.47 & 29.97 & 45.25 & 58.35 & 60.00 \\
& CIF-based model & \textbf{8.02} & \textbf{6.57} & \textbf{7.99} & \textbf{9.73} & \textbf{9.13} \\
\bottomrule
\end{tabular}
}
\vspace{-2mm}
\label{tab:res_long}
\end{table}

We first compare the two models on the long utterances, which are generated by randomly concatenating multiple utterances of the same speaker in the test set, each duration covers at least 50 utterances. For fair comparison, we do not use the long utterance strategy for specific models in \cite{chiu2019comparison, zhou2019improving}.

As shown in table \ref{tab:res_long}, we find the CIF-based model performs more stable on the long utterances, and its deletion and insertion errors are kept at a comparative level for all durations. In contrast, the deletion error of transformer increases rapidly as the length of utterance grows and becomes the main reason of its performance degradation on the long utterances. We also notice the performance gap between the two models on the same durations is different on the two datasets. We suspect it is because the averaged length of training utterances on AISHELL-2 (3.55 seconds) is shorter than Librispeech (12.28 seconds).

Then we compare the two models on the repeated utterances and the noisy utterances. The repeated utterances are generated by randomly selecting 1/10 utterances of each test set, then concatenating the same utterance repeatedly. The number of repetition is 1, 2, 3 and 4 (denoted as $1\times$, $2\times$, $3\times$, $4\times$ in table \ref{tab:res_repeat_noise}). The noisy utterances are generated by mixing every utterance in the test set with one of the 115 noise types in \cite{xu2017multi}, the signal to noise ratio(SNR) ranges from 0-20.

\begin{table}[!ht]
\centering
\vspace{-2mm}
\caption{Results on the repeated and noisy utterances.}
\vspace{-2mm}
\setlength{\tabcolsep}{0.5mm}{
\begin{tabular}{c|c|c|c|c|c||c}
\toprule
Dataset & Model & 1$\times$ & 2$\times$ & 3$\times$ & 4$\times$ & Noisy \\
\midrule
\multirow{2}*{\tabincell{c}{Librispeech \\ (test-clean)}} &
Transformer & \textbf{3.23} & 4.24 & 24.25 & 50.50 & 18.20 \\
& CIF-based model & 3.34 & \textbf{3.48} & \textbf{3.58} & \textbf{3.69} & \textbf{18.15} \\
\midrule
\multirow{2}*{\tabincell{c}{AISHELL-2 \\ (test-ios)}} &
Transformer & 5.73 & 7.12 & 52.7 & 152.92 & 18.95 \\
& CIF-based model & \textbf{5.60} & \textbf{6.20} & \textbf{6.58} & \textbf{6.78} & \textbf{18.89} \\
\bottomrule
\end{tabular}
}
\vspace{-2mm}
\label{tab:res_repeat_noise}
\end{table}

As shown in table \ref{tab:res_repeat_noise}, we find transformer encounters large performance degradation as the number of repetition increases, especially on the $3\times$, $4\times$ setup. Most of errors are insertion errors, which come from lots of cases where more number of repetitions is predicted, there are also some cases where less number of repetition is predicted, but not many. We suspect the degradation is due to the decoding confusion brought by the repeated similar decoder states, which makes transformer hard to predict the label of end of sentence at the correct step. In contrast, the CIF-based model performs stable on the repeated utterances and just encounters slightly performance degradation. Besides, as the repetition times become to $1\times$, $2\times$, $3\times$, we find the inference time becomes to $1\times$, $1.64\times$, $2.63\times$ for the CIF-based model, and becomes to $1\times$, $4.31\times$, $9.88\times$ for transformer on Librispeech, which is basically consistent with the computational complexity for the two models. On the noisy utterances, the two models show similar performance.

The poor generalization of transformer on above cases can be explained from two aspects: 1) transformer uses its decoder state as the query of the encoded representations ($h$) to extract the relevant acoustic information, thus its recognition is affected by both of the decoder state and the $h$, the changes of the two on special cases will bring great challenges to its recognition. In contrast, the integration of acoustic information by CIF-based model is just affected by the $h$. 2) transformer mainly relies on the dependency between the decoder states to determine whether to stop, which is hard to achieve when it suffers from the decoding confusion on some cases. In contrast, the CIF-based model has a clear stop signal that is the last frame is processed. Based on above, we believe the type of frame-synchronous models may be more suitable for the ASR application scenarios that need cover wider range of speech.

\section{Conclusions}
In this work, we make a detailed comparison on a label-synchronous model (transformer) and a soft frame-synchronous model (CIF-based model). Through the experiments on multiple datasets, we find: 1) the label-synchronous model achieves slightly better accuracy on most of datasets, which can be attributed to its comprehensive usage of acoustic information and its insensitivity to the clearness of boundary; 2) the frame-synchronous model achieves 4.4-6.8 times faster inference speed, which is determined by its frame-by-frame calculation and linear computation complexity; 3) the frame-synchronous model also achieves better generalization on the special cases. Since there must be one driving end in the ASR models, we hope the comparison on the two types of models could benefit to the selection and design of end-to-end models in the practical ASR application.

\bibliographystyle{IEEEtran}
\bibliography{mybib}
\end{document}